\documentclass[showpacs,aps,prd,reprint,superscriptaddress,longbibliography]{revtex4-1}
\usepackage[colorlinks=true, pdfstartview=FitV,
bookmarks=true, bookmarksnumbered=true, breaklinks]{hyperref}
\usepackage[dvipdfmx]{graphicx}
\usepackage{amsmath,amssymb,bm,color,longtable,mathrsfs,slashed,comment,tikz}

\definecolor{darkviolet}{rgb}{0.58, 0.0, 0.83}
\definecolor{electricultramarine}{rgb}{0.25, 0.0, 1.0}
\definecolor{brightpink}{rgb}{1.0, 0.0, 0.5}
\hypersetup{linkcolor=brightpink, citecolor=electricultramarine, urlcolor=darkviolet}

\definecolor{lime}{HTML}{A6CE39}
\DeclareRobustCommand{\orcidicon}{
	\hspace{-3mm}
	\begin{tikzpicture}
	\draw[lime, fill=lime] (0,0) 
	circle [radius=0.16] 
	node[white] {{\fontfamily{qag}\selectfont \tiny ID}};
	\draw[white, fill=white] (-0.0625,0.095) 
	circle [radius=0.007];
	\end{tikzpicture}
	\hspace{-3mm}
}

\foreach \x in {A, ..., Z}{\expandafter\xdef\csname orcid\x\endcsname{\noexpand\href{https://orcid.org/\csname orcidauthor\x\endcsname}
			{\noexpand\orcidicon}}
}

\begin{document}
\begin{flushright}
\end{flushright}

\title{Remnants of the nonrelativistic Casimir effect on the lattice}

\author{Katsumasa~Nakayama\orcidA{}}
\email[]{katsumasa.nakayama@riken.jp}
\affiliation{RIKEN Center for Computational Science, Kobe ,650-0047, Japan}
\affiliation{NIC, DESY Zeuthen, Platanenallee 6, 15738 Zeuthen, Germany}

\author{Kei~Suzuki\orcidB{}}
\email[]{k.suzuki.2010@th.phys.titech.ac.jp}
\affiliation{Advanced Science Research Center, Japan Atomic Energy Agency (JAEA), Tokai 319-1195, Japan}

\date{\today}

\begin{abstract}
The Casimir effect is a fundamental quantum phenomenon induced by the zero-point energy for a quantum field.
It is well-known for relativistic fields with a linear dispersion relation, while its existence or absence for nonrelativistic fields with a quadratic dispersion is an unsettled question.
Here, we investigate the Casimir effects for various dispersion relations on the lattice.
We find that Casimir effects for dispersions proportional to an even power of momentum are absent in a long distance but a remnant of the Casimir effect survives in a short distance.
Such a remnant Casimir effect will be experimentally observed in materials with quantum fields on the lattice, such as thin films, narrow nanoribbons, and short nanowires.
In terms of this effect, we also give a reinterpretation of the Casimir effect for massive fields.
\end{abstract}

\maketitle

\section{Introduction} \label{Sec:1}
Zero-point energy is a fundamental concept predicted in quantum mechanics. 
In fact, the Casimir effect \cite{Casimir:1948dh} is a quantum phenomenon induced by the zero-point energy for photons confined by a spatial boundary condition and was experimentally established ~\cite{Lamoreaux:1996wh} (see Refs.~\cite{Plunien:1986ca,Mostepanenko:1988bs,Bordag:2001qi,Milton:2001yy,Klimchitskaya:2009cw} for reviews).
The photonic Casimir effect will be useful in the field of nanophotonics~\cite{Gong:2020ttb}, while analogous effects for quantum fields on the lattice in solid states, such as electrons, phonons, and magnons, will open engineering fields which may be named {\it Casimir electronics} and {\it Casimir spintronics}.

\begin{figure}[tb!]
    \centering
    \begin{minipage}[t]{0.5\columnwidth}
    \includegraphics[clip,width=1.0\columnwidth]{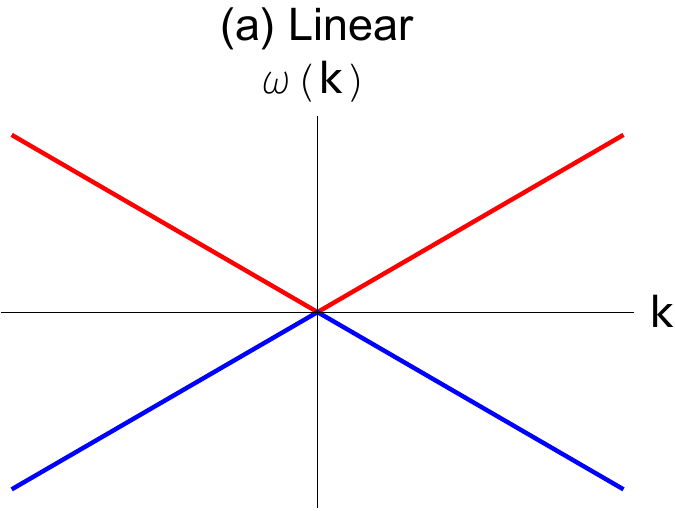}
    \end{minipage}%
    \begin{minipage}[t]{0.5\columnwidth}
    \includegraphics[clip,width=1.0\columnwidth]{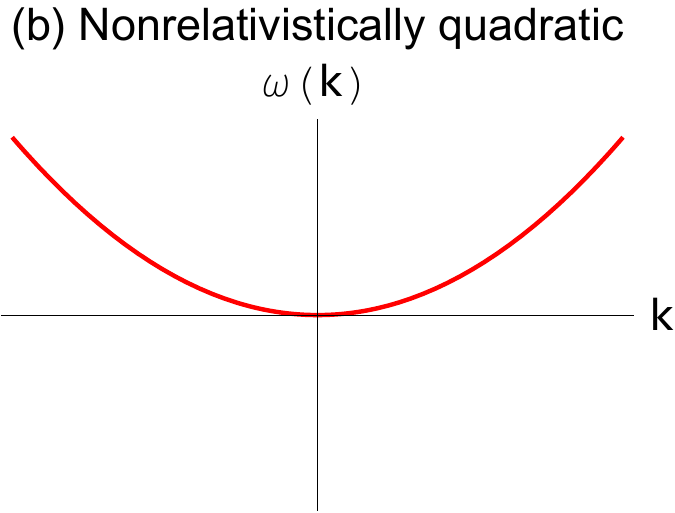}
    \end{minipage}
    \begin{minipage}[t]{0.5\columnwidth}
    \includegraphics[clip,width=1.0\columnwidth]{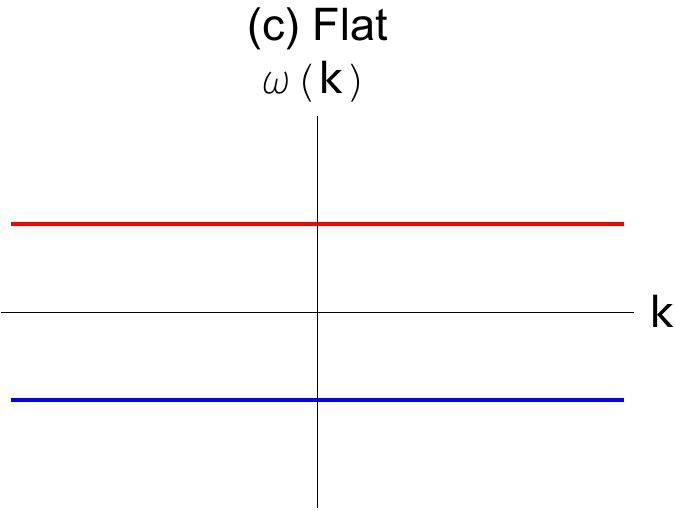}
    \end{minipage}%
    \begin{minipage}[t]{0.5\columnwidth}
    \includegraphics[clip,width=1.0\columnwidth]{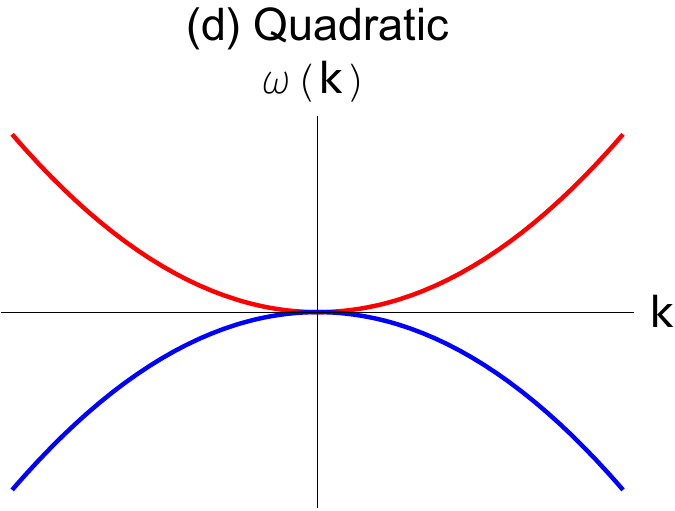}
    \end{minipage}    
    \caption{Examples of typical dispersion relations.
(a) A relativistic linear dispersion. (b) A nonrelativistically quadratic dispersion. (c) A flat dispersion. (d) A quadratic dispersion.}
    \label{disp}
\end{figure}

For relativistic degrees of freedom with the linear dispersion relations as shown in Fig.~\ref{disp}(a), such as the photon and the massless Dirac fermion, the occurrence of the Casimir effect is well known.
On the other hand, an unsettled question is whether the Casimir effect for nonrelativistic fields exists.
In general, the existence or absence of the Casimir effect for a quantum field depends on the following three points:
\begin{enumerate}
\item[] (i) the existence/absence of zero-point energy,
\item[] (ii) the form of the dispersion relation, and
\item[] (iii) the spatial boundary condition.
\end{enumerate}
Absence of the Casimir effect can be realized by any of these conditions.
For example, when we consider a free Schr\"{o}dinger field with a quadratic dispersion as shown in Fig.~\ref{disp}(b), as usual in solid state physics, we can prove that the zero-point energy is exactly zero.
In this sense, by reason (i), we may say that the Casimir effect for nonrelativistic fields described as the Schr\"{o}dinger field does not occur.
For reason (ii), there are some examples.
As shown in Fig.~\ref{disp}(c), a flat band has no momentum-dependent dispersion.
For such a constant dispersion, the discretization of momenta in a finite volume never produces any difference from the zero-point energy in the infinite volume, so that the Casimir effect does not occur.
As another example, as shown in Fig.~\ref{disp}(d), for a field with two quadratic dispersions consisting of positive and negative eigenvalues, one can prove that the Casimir effect disappears under a boundary condition (see, e.g., Refs.~\cite{Cougo-Pinto:2001,Fulling:2003zx,Kolomeisky:2013zra,Ulhoa:2017tsg} and a textbook~\cite{Milton:2001yy})
\footnote{
Using the zeta function regularization and the dimensional regularization, the Casimir energy for a dispersion relation with the order of $s$ in the $d+1$ dimensional spacetime and the periodic boundary condition is represented as
\begin{equation}
    E_\mathrm{Cas}^{[s]}
    =2 \frac{1}{(4\pi)^{(d-1)/2}}\frac{\pi^{-1/2}\Gamma\left( \frac{d+s}{2} \right) \zeta(d+s)}{\Gamma(-\frac{s}{2})}
    \frac{2^{d-1+s}}{L^{d-1+s}}. 
\end{equation}
When $s$ is an even number, $E_\mathrm{Cas}^{[s]}=0$ because of $\Gamma(-\frac{s}{2})$.
Also, an alternative interpretation, it is well known that the Casimir energy for relativistic fields with a nonzero and finite mass is characterized by the modified Bessel function, and its infinite-mass limit goes to zero~\cite{Hays:1979bc,Ambjorn:1981xw,Plunien:1986ca}.}.
Also, in this sense, the Casimir effect for nonrelativistic fields seems to be prohibited.
However, the understanding of such a tendency is still not clear.

In this letter, we theoretically study the Casimir effect for degrees of freedom with some types of dispersion relations on the lattice, by using a lattice regularization approach \cite{Ishikawa:2020ezm,Ishikawa:2020icy}\footnote{Studies of the Casimir effect using the lattice regularization are still few.
For early works for scalar fields, see Refs.~\cite{Actor:1999nb,Pawellek:2013sda}.
}.
We show that the Casimir effect for quantum fields on the lattice with an even-order dispersion tends to be absent (which is consistent with the case in continuous space), but a remnant of the Casimir effect survives in a short distance (which is distinct from the continuous case).
We call this phenomenon the {\it remnant Casimir effect}.
This effect gives us a deeper understanding of the Casimir effect, and moreover, it will be experimentally observed in finite-size condensed matter systems, such as thin films, narrow nanoribbons, and short nanowires of materials.
Throughout this letter, we use the natural units $\hbar = v = 1$ for the 
reduced Planck constant $\hbar$ and the velocity of a particle $v$.

\section{Casimir effect on the lattice}
First, we define the Casimir effect on the lattice.
For simplicity, we consider a real scalar field in the three-dimensional (3D) lattice space (the cases in other spatial dimensions are straightforward).
We set the periodic boundary condition (other boundaries will be discussed later) for the $z$ axis in momentum space as $k_z \to 2l \pi /L_z$ or $a k_z \to 2l\pi/N_z$, where $L_z\equiv aN_z$ is the distance with the lattice constant $a$, and $l=0,1,\cdots, N_z-1$ within the first Brillouin zone (BZ), $0 \leq ak_i < 2\pi$ ($i=x,y,z$).
The Casimir energy $E_\mathrm{Cas}$ for the lattice number $N_z$ on the $z$ axis (per the lattice number $N_xN_y$ on the surface area) is defined as the difference between the zero-point energies $E_0^\mathrm{sum}$ with discrete energy levels $\omega_{k_\perp,l}$ and $E_0^\mathrm{int}$ with a continuous energy level $\omega_{\bf k}$~\cite{Ishikawa:2020ezm,Ishikawa:2020icy} (also see the Supplemental Material~\footnote{In Supplemental Material S1, we compare the definitions of Casimir energies in the continuum space and in the lattice space.
In S2, we show the detailed analyses of the phononic Casimir effect in a one-dimensional GaAs nanowire.}):
\begin{align}
aE_\mathrm{Cas} &\equiv aE_0^\mathrm{sum}(N_z) - aE_0^\mathrm{int}(N_z), \label{eq:def_cas} \\
aE_0^\mathrm{sum}(N_z) &= \int_\mathrm{BZ} \frac{d^2 (ak_\perp)}{(2\pi)^2} \left[ \frac{1}{2}  \sum_{l=0}^{N_z-1} a|\omega_{k_\perp,l}|  \right], \label{eq:def_cas_disc}  \\
aE_0^\mathrm{int}(N_z) & = \int_\mathrm{BZ} \frac{d^2(ak_\perp)}{(2\pi)^2} \left[ \frac{N_z}{2} \int_\mathrm{BZ} \frac{d(ak_z)}{2\pi} a|\omega_{\bf k}| \right], \label{eq:def_cas_cont}
\end{align}
where the momentum integral is over the first BZ, $k_\perp^2 \equiv k_x^2+k_y^2$, and $d^2(ak_\perp) \equiv d(ak_x)d(ak_y)$.
The factor of $+\frac{1}{2}$ comes from the zero-point energy for the real scalar field.

\section{Dispersion relation dependence of Casimir effect}
By using the definition in Eq.~(\ref{eq:def_cas}), we can investigate the Casimir energies for various dispersion relations.
Massless dispersion relations with a 3D momentum ${\bf k} =(k_x,k_y,k_z)$ in continuous space are defined as
\begin{align}
&a\omega_{\bf k}= \pm |a{\bf k}|^s = \pm (a^2 k_x^2+a^2 k_y^2+a^2k_z^2)^{s/2},
\end{align}
where $s$ is the order of dispersion.
For example, $s=1$ corresponds to a field with the linear dispersion $\omega_{\bf k}= \pm |{\bf k}|$, such as the massless Klein-Gordon field.
Here, $s=2$ is a field with the quadratic dispersion $\omega_{\bf k}= \pm {\bf k}^2$.
A dispersion relation in lattice space is obtained by replacing $a{\bf k}$ with ${a\tilde{\bf k}} =a(\tilde{k}_x,\tilde{k}_y,\tilde{k}_z)$ using $a^2\tilde{k}_i^2 = 2-2\cos a k_i$:
\begin{align}
&a\omega_{\bf k}= \pm |a\tilde{{\bf k}}|^s = \pm [\sum_{i=x,y,z} (2-2\cos ak_i) ]^{s/2}. \label{omega_lat}
\end{align}
By substituting this form into Eqs.~(\ref{eq:def_cas_disc}) and (\ref{eq:def_cas_cont}), we can calculate the Casimir energy.

\begin{figure}[tb!]
    \centering
    \begin{minipage}[t]{0.84\columnwidth}
    \includegraphics[clip,width=1.0\columnwidth]{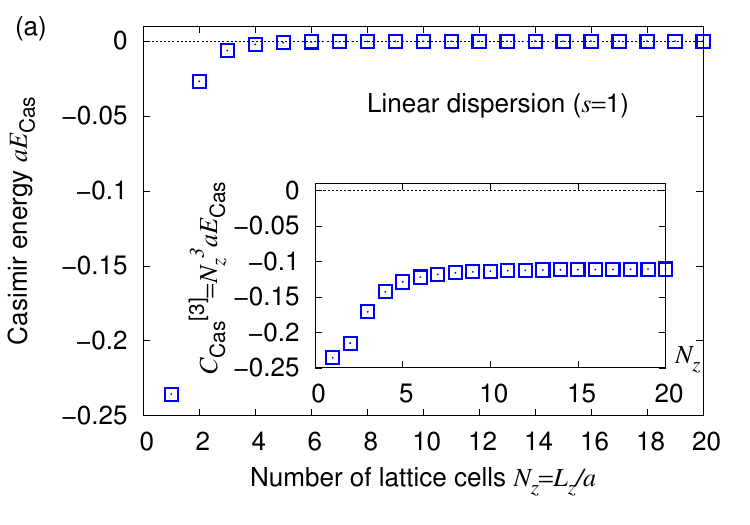}
    \end{minipage}
    \begin{minipage}[t]{0.84\columnwidth}
    \includegraphics[clip,width=1.0\columnwidth]{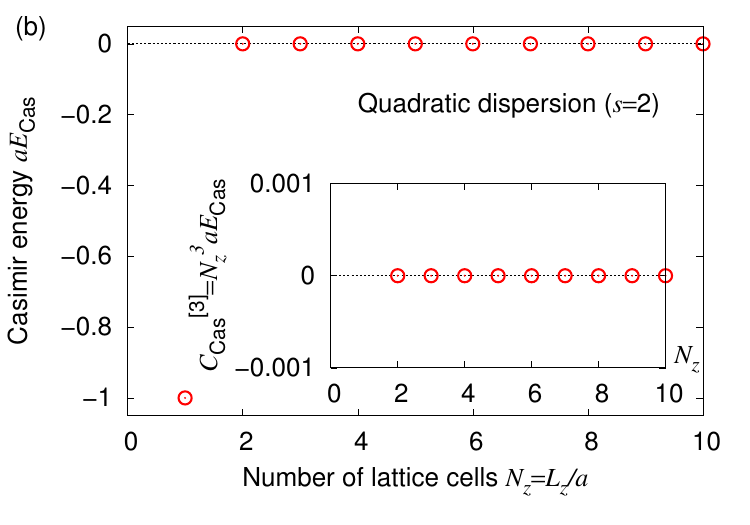}
    \end{minipage}
    \begin{minipage}[t]{0.84\columnwidth}
    \includegraphics[clip,width=1.0\columnwidth]{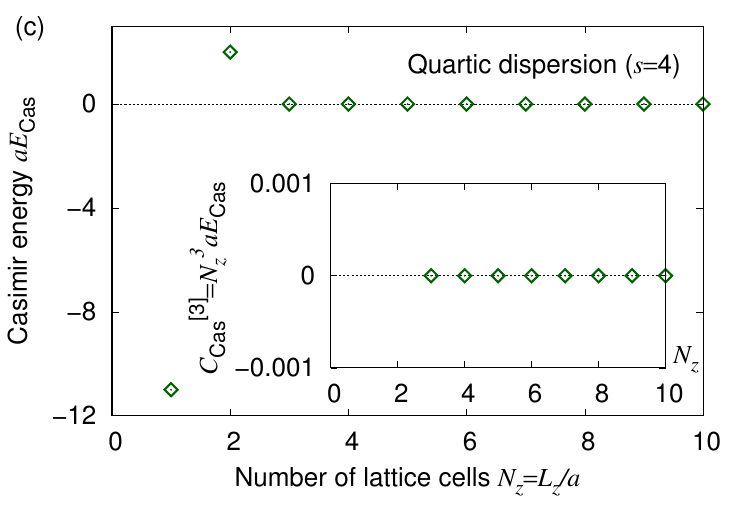}
    \end{minipage}
    \begin{minipage}[t]{0.84\columnwidth}
    \includegraphics[clip,width=1.0\columnwidth]{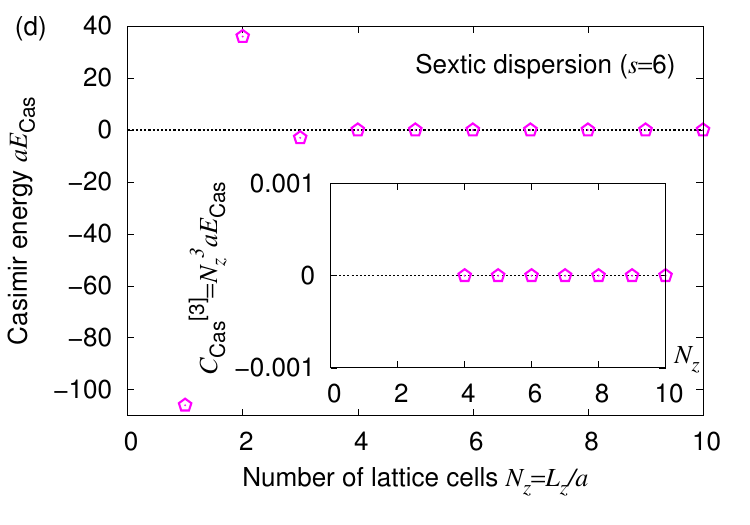}
    \end{minipage}
    \caption{Casimir energies $aE_\mathrm{Cas}$ for massless real scalar fields on the lattice, where the dispersion relation is (a) linear ($s=1$), (b) quadratic ($s=2$), (c) quartic ($s=4$), or  (d) sextic  ($s=6$) order.
Insets: Casimir coefficients $C_\mathrm{Cas}^{[3]}$.
The behaviors for $s=$ 2, 4, and $6$ are the remnant Casimir effect.}
    \label{Ecas}
\end{figure}

In Fig.~\ref{Ecas}, we show the results for the dimensionless Casimir energy $aE_\mathrm{Cas}$ and the Casimir coefficient (which is a convenient quantity to check the $N_z^3$ dependence of $aE_\mathrm{Cas}$) $C_\mathrm{Cas}^{[3]} \equiv N_z^3aE_\mathrm{Cas}$ for the linear ($s=1$), quadratic ($s=2$), quartic ($s=4$), or sextic ($s=6$) dispersion relation.
The result at $s=1$ is evaluated by numerical integrals, and the others can be analytically obtained. 
For $s=1$, we see the well-known behaviors of the Casimir effect: $aE_\mathrm{Cas}$ decreases as $N_z$ increases.
In the large $N_z$ region, $C_\mathrm{Cas}^{[3]}$ approximately approaches the analytic solution known in continuous space $C_\mathrm{Cas}^{[3]}=-\pi^2/90$.
On the other hand, the result for $s=2$ is quite distinct, where we find that a nonzero $E_\mathrm{Cas}$ survives only at $N_z=1$, while $E_\mathrm{Cas}$ is exactly zero at $N_z>1$.
It is easy to analytically derive this behavior: using $aE_0^\mathrm{sum} = 2$ and $aE_0^\mathrm{int} = 3$ at $N_z=1$, we obtain $aE_\mathrm{Cas} \equiv aE_0^\mathrm{sum} -aE_0^\mathrm{int}= -1$, while using $aE_0^\mathrm{sum} = aE_0^\mathrm{int} $ at $N_z>1$, $E_\mathrm{Cas}=0$.
Furthermore, we can easily derive that this behavior does not depend on the spatial dimension.
Similarly, also for $s=2n$ ($n=1,2,\cdots$), we find a nonzero $E_\mathrm{Cas}$ only at $N_z \leq n$.
We call this behavior the remnant Casimir effect.
Note that, for odd-order dispersions $s=2n-1$,  $E_\mathrm{Cas}$ is nonzero at any $N_z$, and $C_\mathrm{Cas}^{[2+s]} \equiv N_z^{2+s}aE_\mathrm{Cas}$ approximately approaches a constant value.

We comment on the other types of boundary conditions.
For the antiperiodic boundary [$ak_z \to (2l+1) \pi /N_z$ with $l=0,\cdots,N_z-1$], the remnant Casimir effect does not change qualitatively, except for its magnitude and sign.
For a phenomenological boundary $[ak_z \to l \pi /N_z$ with $l=1,\cdots,2N_z$ (or $l=0,\cdots,2N_z-1)$], there are no remnants at any $N_z$ for $s=2$, while we find a remnant appears at $N_z=1$ for $s=4,6$ and remnants at $N_z< n+1$ for $s=4n,4n+2$.
Thus, various boundary conditions can induce the remnant Casimir effect.

\begin{table}[t!]
  \centering
  \caption{Classification of Casimir effects for fields with various dispersion relations in the continuous spacetime or the lattice space defined by Eq.~(\ref{omega_lat}).}
\begin{tabular}{c l l } 
  \hline \hline
  Order $s$ & Continuous & Lattice (periodic boundary) \\ 
  \hline
  $1(m=0)$  & Lasting & Lasting \\
  $1(m\neq0, m \neq \infty)$  & Damping & Damping \\
  $2$  & No & Remnant ($N_z=1$) \\
  $4$  & No & Remnant ($N_z=1,2$) \\
  $\vdots$ &  & \\
  $2n$ &No & Remnant ($N_z=1,2,\cdots,n$) \\
  $2n-1(m=0)$    & Lasting & Lasting \\
  \hline \hline   
  \end{tabular}
  \label{tab:kinds}
\end{table}

\section{Classification of Casimir effects}
Based on the above investigation, we classify typical behaviors of Casimir effects into four types (see Table~\ref{tab:kinds} for the periodic boundary):
\begin{itemize}
\item No Casimir effect---In any distance, the Casimir energy is exactly zero.
A representative example is a field with the even-order dispersion in continuous space.
\item Lasting Casimir effect---The Casimir energy is nonzero in a long distance, and the Casimir coefficient approaches asymptotically to a nonzero value.
Such a behavior is known for fields with a linear dispersion in both the continuous and lattice space.
\item Damping Casimir effect---The Casimir energy is nonzero in a long distance, but the Casimir coefficient asymptotically approaches zero.
Such a behavior is well known for massive fields~\cite{Hays:1979bc,Ambjorn:1981xw,Plunien:1986ca} because a mass parameter usually tends to suppress the Casimir effect in a long distance.
\item Remnant Casimir effect---The Casimir energy in a long distance is exactly zero, but nonzero values (remnants) survive only in a short distance.
Such a behavior is realized in fields with an even-order dispersion in lattice space.
This is the main finding in this letter. 
\end{itemize}

\begin{figure}[t!]
    \centering
    \begin{minipage}[t]{0.5\columnwidth}
    \includegraphics[clip,width=1.0\columnwidth]{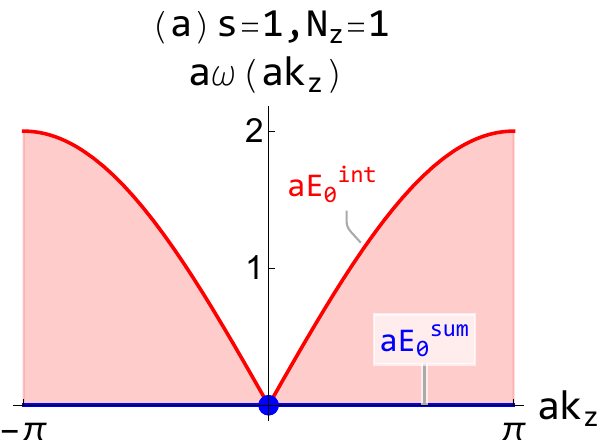}
    \end{minipage}%
    \begin{minipage}[t]{0.5\columnwidth}
    \includegraphics[clip,width=1.0\columnwidth]{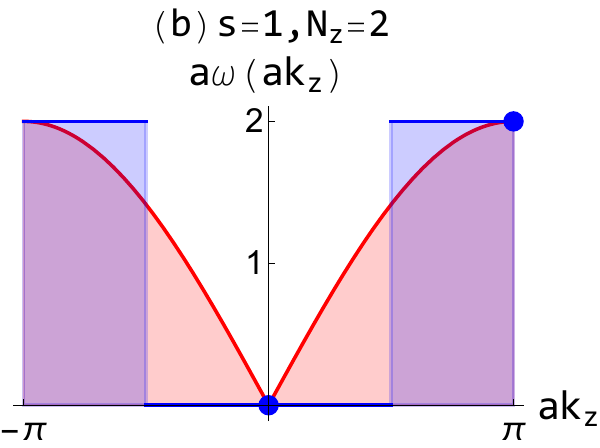}
    \end{minipage}
    \begin{minipage}[t]{0.5\columnwidth}
    \includegraphics[clip,width=1.0\columnwidth]{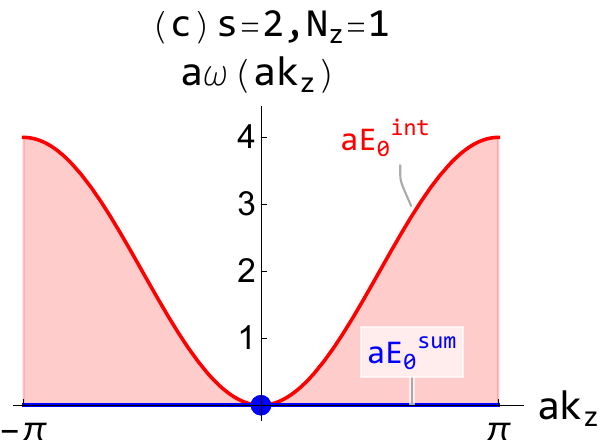}
    \end{minipage}%
    \begin{minipage}[t]{0.5\columnwidth}
    \includegraphics[clip,width=1.0\columnwidth]{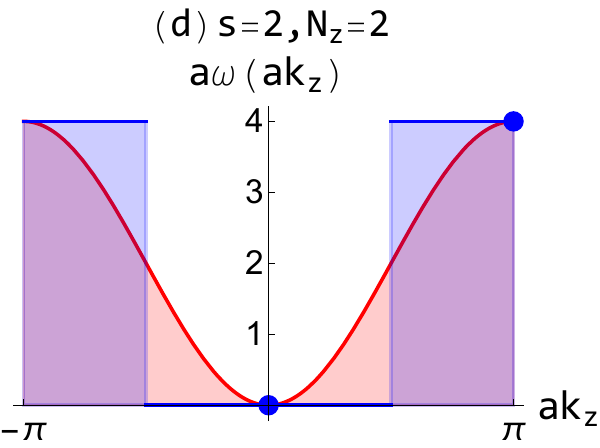}
    \end{minipage}
    \begin{minipage}[t]{0.5\columnwidth}
    \includegraphics[clip,width=1.0\columnwidth]{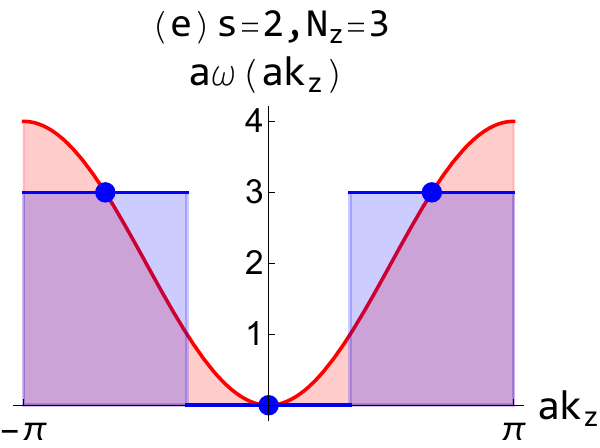}
    \end{minipage}%
    \begin{minipage}[t]{0.5\columnwidth}
    \includegraphics[clip,width=1.0\columnwidth]{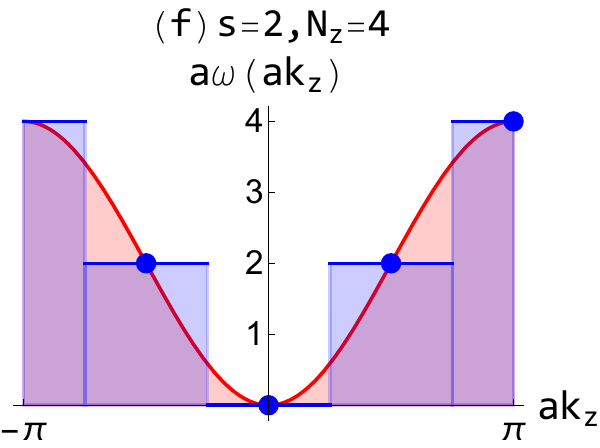}
    \end{minipage}   
    \caption{Graphical interpretation for Casimir effects on the lattice.
Blue points: discrete energy levels.
Red curves: continuous energy levels.
Blue and red colored regions correspond to $aE_0^\mathrm{sum}$ and $aE_0^\mathrm{int}$, respectively.
(a) and (b) Linear dispersion at $N_z=1,2$.
(c)--(f) Quadratic dispersion at $N_z=1,2,3,4$.}
    \label{rectangle}
\end{figure}

\section{Graphical interpretation}
Here, we give a graphical interpretation for Casimir effects on the lattice.
By the definition in Eq.~(\ref{eq:def_cas}), the Casimir energy is defined as the difference between $E_0^\mathrm{sum}$ and $E_0^\mathrm{int}$, where $E_0^\mathrm{sum}$ is the sum over discretized $k_z$, and $E_0^\mathrm{int}$ is the integral with respect to continuous $k_z$.
As shown in Fig.~\ref{rectangle}, such a difference can be graphically understood by comparing (i) the sum of the areas of rectangles with a width $2\pi/N_z$ and a height $[2-2\cos (2l\pi /N_z)]^{s/2}$ (blue region) and (ii) the integral of the dispersion relation with respect to $ak_z$ (red region), within a BZ.
For example, Figs.~\ref{rectangle}(a) and (b) for $s=1$ show that $E_0^\mathrm{sum}$ and $E_0^\mathrm{int}$ are different from each other.
Such a difference survives at any $N_z$, so that the resultant Casimir energies are nonzero at any $N_z$.
Similarly, in Fig.~\ref{rectangle}(c) for $s=2$ at $N_z=1$, this situation does not change, which is the origin of the only remnant for $s=2$.
On the other hand, in Figs.~\ref{rectangle}(d)--\ref{rectangle}(f) for $s=2$ at $N_z > 1$, we find an exact cancellation between $E_0^\mathrm{sum}$ and $E_0^\mathrm{int}$, which leads to the exactly zero Casimir energy.
Mathematically, this is nothing but the property of $2-2\cos ak_z$.
Thus, the Casimir effect on the lattice can be visually understood as a difference or cancellation of the zero-point energies.

\section{Reinterpretation of massive dispersion}
Here, we discuss a connection between the the Casimir effect for the $s=1$ massive dispersion and the remnant Casimir effect for even-order dispersions.
The $s=1$ dispersion relation with a mass (or a gap) $m$ is expanded around $|{\bf k}|/m = 0$ as
\begin{align}
\omega_{\bf k} &= \pm \sqrt{{\bf k}^2 +m^2} \nonumber\\
&= \pm \left( m +\frac{ {\bf k}^2}{2m} -\frac{ {\bf k}^4}{8m^3} + \frac{ {\bf k}^6}{16m^5} + \cdots \right).
\end{align}
The first term is a flat band without any momentum dependence, as in Fig.~\ref{disp}(c), and never contributes to the Casimir effect.
The other terms are regarded as the sum of various even-order dispersion relations.

\begin{figure}[t!]
    \centering
    \begin{minipage}[t]{1.0\columnwidth}
    \includegraphics[clip,width=1.0\columnwidth]{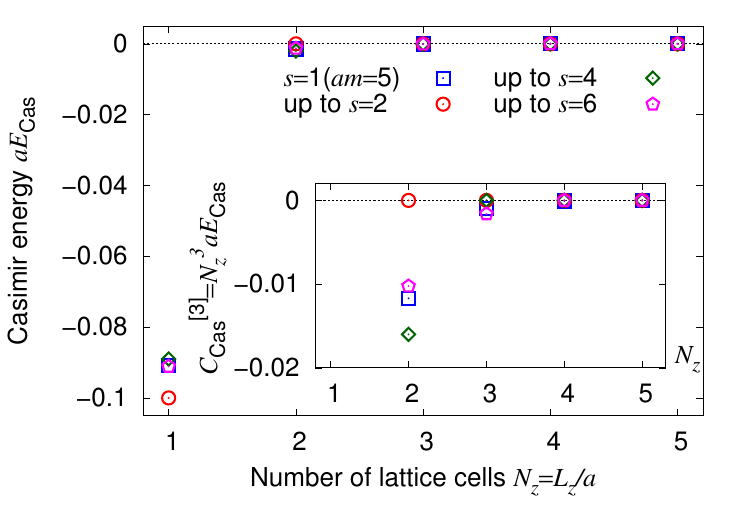}
    \end{minipage}
    \caption{Comparison between the Casimir energy for a massive real scalar field and the sum of remnant Casimir energies for lower-even-order dispersions.
The mass parameter is fixed as $am=5$.}
    \label{expa}
\end{figure}

Using this form, the zero-point energy is rewritten as
\begin{align}
E_0& = \int \frac{d{\bf k}}{(2\pi)^3} \, \omega_{\bf k} = \int \frac{d{\bf k}}{(2\pi)^3} \, \sqrt{{\bf k}^2 +m^2}  \nonumber\\
&= \int \frac{d{\bf k}}{(2\pi)^3} \, \left( m +\frac{ {\bf k}^2}{2m} -\frac{ {\bf k}^4}{8m^3} + \frac{ {\bf k}^6}{16m^5} + \cdots \right). \label{eq:E0_exp}
\end{align}
We reexamine the Casimir effect from the last form of the zero-point energy by using lattice regularization~\footnote{By using the dispersion relation $a\tilde{\bf k}$ on the lattice, the Taylor expansion of Eq.~(\ref{eq:E0_exp}) converges when $\sum_{i} (2-2\cos ak_i)/(am)^2<1$}.
In Fig.~\ref{expa}, we show the comparison between the Casimir energies for the massive dispersion calculated from the third form of Eq.~(\ref{eq:E0_exp}) and the sum of remnant Casimir energies for lower even-order dispersion ($s=2,4,6, \cdots$) calculated from the last form of Eq.~(\ref{eq:E0_exp}), where we fix the dimensionless mass as $am=5$.
Even at $N_z=1$, the sum of the remnants up to $s=4$ is almost consistent with that for the massive dispersion.
We find that, even at $N_z=3$, the remnants up to $s=6$ become a good approximation.
Thus, we have derived the picture that the Casimir energy for $s=1$ massive fields on the lattice is regarded as the sum of the remnant Casimir energies for massless fields of lower even orders.

It should be noted that, in continuous space, this picture of finite sum is not correct since the Casimir energies for even-order dispersion relations are exactly zero.
The Casimir energy in continuous space is obtained only by infinite summation of all orders.
This inconsistency is because the dispersion relation on the lattice is slightly different from that in continuous space.

\section{Experimental realizations}
We emphasize that not only the lattice Casimir effect but also the remnant Casimir effect intrinsically arise in relativistic condensed matter systems when its energy levels are discretized by the finite size of a system.
For example, a quadratic dispersion relation as shown in Fig.~\ref{disp}(d) appears in AB-stacked bilayer graphene~\cite{McCann:2005xf,McCann:2013}, multi-Weyl semimetals~\cite{Xu:2011dn,Fang:2012,Huang:2016}, and AB-stacked honeycomb bilayer magnets~\cite{Owerre:2016}.
In such materials, the Casimir effect is intrinsically realized on nanoribbons of two-dimensional materials and in thin films of 3D materials, and the Casimir energy is defined as an energy difference relative to the infinite bulk.
In this sense, the Casimir energy is a physical quantity well defined in principle, and thus, the remnant Casimir effect in even-order dispersion systems is also well defined.
We stress that such an energy difference can influence thermodynamic properties of materials, such as magnetization.
As a result, the Casimir effect can be measured: One can observe it as a remnant behavior of a physical quantity in a short distance and an unchanged behavior in a long distance.

We remark on two points.
First, the magnitude of the remnant Casimir energy in a short distance is generally comparable with the free energy of the original (infinite-bulk) system (which is proportional to $aE_0^\mathrm{int}$) because this remnant is generated from a large difference between $aE_0^\mathrm{sum}$ and $aE_0^\mathrm{int}$.
Second, the unchanged behavior in a long distance, the disappearance of the Casimir effect, can be regarded as a no-go theorem of the Casimir effect, which is always helpful for the correct understanding for finite-size effects.
Note that the band structures in realistic materials may be slightly different from the exact $2-2\cos{ak}$ band due to the existence of interactions or impurities, but even in such a case, one can observe an approximate effect: approximate remnants in a short distance and approximate zeros in a long distance.
In addition, if one knows the form of $2-2\cos{ak}$ in the infrared region but does not know its ultraviolet structure, the nonzero Casimir effect in a long distance implies a difference from $2-2\cos{ak}$ in the ultraviolet region.

\begin{figure}[t!]
    \centering
    \begin{minipage}[t]{1.0\columnwidth}
    \includegraphics[clip,width=1.0\columnwidth]{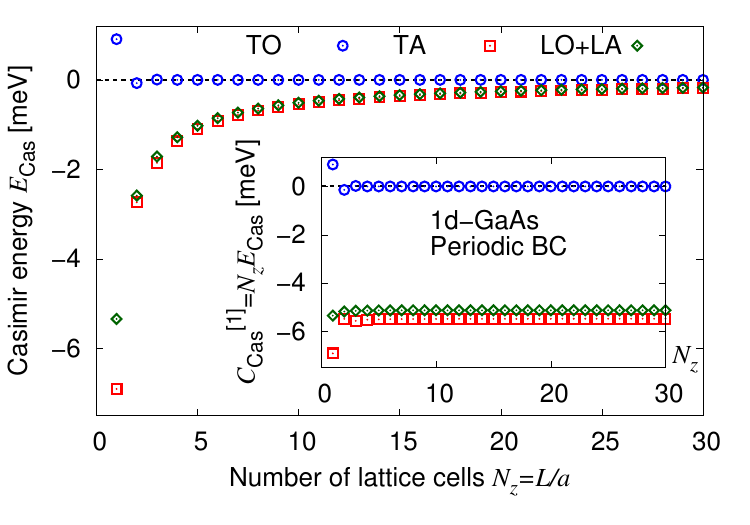}
    \end{minipage}
    \caption{Casimir energy and Casimir coefficient for phonons in one-dimensional-like GaAs nanowires with the periodic boundary condition.
}
    \label{fig:GaAs_PBC}
\end{figure}

For example, we consider phonon fields in GaAs nanowires along the $z$ axis, where there are four types of phonons: the transverse optical (TO), longitudinal optical (LO), transverse acoustic (TA), and longitudinal acoustic (LA) modes.
For dispersion relations, we use a model with parameters estimated by fitting the experimental data of the four modes in the whole BZ~\cite{Strauch:1990} (See Supplemental Material).
In Fig.~\ref{fig:GaAs_PBC}, we show the numerical results for the Casimir energy with the periodic boundary.
We find that $E_\mathrm{Cas}$ for the TO modes is strongly suppressed compared with the other modes, except for $N_z=1$.
This is because the dispersion relations for the TO modes are quadraticlike around both $ak_z=0$ and $\pi$, and other modes are not. 
This is regarded as an approximate realization of the remnant Casimir effect.

\section{Conclusions}
In this letter, we have proposed the remnant Casimir effect which is a unique property of even-order dispersion on the lattice.
This is a well-defined concept in principle and should be experimentally realized.
Our findings will become common knowledge on Casimir effects on the lattice and will play a key role in the Casimir engineering, which is the application of quantum fluctuations on the lattice to engineering fields, such as Casimir photonics~\cite{Gong:2020ttb}, Casimir electronics, and Casimir spintronics.

\section*{ACKNOWLEDGMENTS}
The authors thank Yasufumi Araki and Kouki Nakata for fruitful discussions.
This letter was supported by the Japan Society for the Promotion of Science (JSPS) KAKENHI (Grants No. JP17K14277 and No. JP20K14476).

\bibliography{ref}

\begin{thebibliography}{30}%
\makeatletter
\providecommand \@ifxundefined [1]{%
 \@ifx{#1\undefined}
}%
\providecommand \@ifnum [1]{%
 \ifnum #1\expandafter \@firstoftwo
 \else \expandafter \@secondoftwo
 \fi
}%
\providecommand \@ifx [1]{%
 \ifx #1\expandafter \@firstoftwo
 \else \expandafter \@secondoftwo
 \fi
}%
\providecommand \natexlab [1]{#1}%
\providecommand \enquote  [1]{``#1''}%
\providecommand \bibnamefont  [1]{#1}%
\providecommand \bibfnamefont [1]{#1}%
\providecommand \citenamefont [1]{#1}%
\providecommand \href@noop [0]{\@secondoftwo}%
\providecommand \href [0]{\begingroup \@sanitize@url \@href}%
\providecommand \@href[1]{\@@startlink{#1}\@@href}%
\providecommand \@@href[1]{\endgroup#1\@@endlink}%
\providecommand \@sanitize@url [0]{\catcode `\\12\catcode `\$12\catcode
  `\&12\catcode `\#12\catcode `\^12\catcode `\_12\catcode `\%12\relax}%
\providecommand \@@startlink[1]{}%
\providecommand \@@endlink[0]{}%
\providecommand \url  [0]{\begingroup\@sanitize@url \@url }%
\providecommand \@url [1]{\endgroup\@href {#1}{\urlprefix }}%
\providecommand \urlprefix  [0]{URL }%
\providecommand \Eprint [0]{\href }%
\providecommand \doibase [0]{http://dx.doi.org/}%
\providecommand \selectlanguage [0]{\@gobble}%
\providecommand \bibinfo  [0]{\@secondoftwo}%
\providecommand \bibfield  [0]{\@secondoftwo}%
\providecommand \translation [1]{[#1]}%
\providecommand \BibitemOpen [0]{}%
\providecommand \bibitemStop [0]{}%
\providecommand \bibitemNoStop [0]{.\EOS\space}%
\providecommand \EOS [0]{\spacefactor3000\relax}%
\providecommand \BibitemShut  [1]{\csname bibitem#1\endcsname}%
\let\auto@bib@innerbib\@empty
\bibitem [{\citenamefont {Casimir}(1948)}]{Casimir:1948dh}%
  \BibitemOpen
  \bibfield  {author} {\bibinfo {author} {\bibfnamefont {H.~B.~G.}\
  \bibnamefont {Casimir}},\ }\bibfield  {title} {\enquote {\bibinfo {title}
  {{On the Attraction Between Two Perfectly Conducting Plates}},}\ }\href@noop
  {} {\bibfield  {journal} {\bibinfo  {journal} {Proc. Kon. Ned. Akad.
  Wetensch.}\ }\textbf {\bibinfo {volume} {51}},\ \bibinfo {pages} {793}
  (\bibinfo {year} {1948})}\BibitemShut {NoStop}%
\bibitem [{\citenamefont {Lamoreaux}(1997)}]{Lamoreaux:1996wh}%
  \BibitemOpen
  \bibfield  {author} {\bibinfo {author} {\bibfnamefont {S.~K.}\ \bibnamefont
  {Lamoreaux}},\ }\bibfield  {title} {\enquote {\bibinfo {title}
  {{Demonstration of the Casimir Force in the 0.6 to $6\ensuremath{\mu}m$
  Range}},}\ }\href {https://doi.org/10.1103/PhysRevLett.78.5} {\bibfield
  {journal} {\bibinfo  {journal} {Phys. Rev. Lett.}\ }\textbf {\bibinfo
  {volume} {78}},\ \bibinfo {pages} {5} (\bibinfo {year} {1997})},\ \bibinfo
  {note} {[Erratum: \href{https://doi.org/10.1103/PhysRevLett.81.5475}{Phys.
  Rev. Lett. {\bf 81}, 5475(E) (1998)}]}\BibitemShut {NoStop}%
\bibitem [{\citenamefont {Plunien}\ \emph {et~al.}(1986)\citenamefont
  {Plunien}, \citenamefont {M\"uller},\ and\ \citenamefont
  {Greiner}}]{Plunien:1986ca}%
  \BibitemOpen
  \bibfield  {author} {\bibinfo {author} {\bibfnamefont {G.}~\bibnamefont
  {Plunien}}, \bibinfo {author} {\bibfnamefont {Berndt}\ \bibnamefont
  {M\"uller}}, \ and\ \bibinfo {author} {\bibfnamefont {W.}~\bibnamefont
  {Greiner}},\ }\bibfield  {title} {\enquote {\bibinfo {title} {{The Casimir
  effect}},}\ }\href {https://doi.org/10.1016/0370-1573(86)90020-7} {\bibfield
  {journal} {\bibinfo  {journal} {Phys. Rept.}\ }\textbf {\bibinfo {volume}
  {134}},\ \bibinfo {pages} {87} (\bibinfo {year} {1986})}\BibitemShut
  {NoStop}%
\bibitem [{\citenamefont {Mostepanenko}\ and\ \citenamefont
  {Trunov}(1988)}]{Mostepanenko:1988bs}%
  \BibitemOpen
  \bibfield  {author} {\bibinfo {author} {\bibfnamefont {V.~M.}\ \bibnamefont
  {Mostepanenko}}\ and\ \bibinfo {author} {\bibfnamefont {N.~N.}\ \bibnamefont
  {Trunov}},\ }\bibfield  {title} {\enquote {\bibinfo {title} {{The Casimir
  effect and its applications}},}\ }\href
  {https://doi.org/10.1070/PU1988v031n11ABEH005641} {\bibfield  {journal}
  {\bibinfo  {journal} {Sov. Phys. Usp.}\ }\textbf {\bibinfo {volume} {31}},\
  \bibinfo {pages} {965} (\bibinfo {year} {1988})}\BibitemShut {NoStop}%
\bibitem [{\citenamefont {Bordag}\ \emph {et~al.}(2001)\citenamefont {Bordag},
  \citenamefont {Mohideen},\ and\ \citenamefont
  {Mostepanenko}}]{Bordag:2001qi}%
  \BibitemOpen
  \bibfield  {author} {\bibinfo {author} {\bibfnamefont {Michael}\ \bibnamefont
  {Bordag}}, \bibinfo {author} {\bibfnamefont {U.}~\bibnamefont {Mohideen}}, \
  and\ \bibinfo {author} {\bibfnamefont {V.~M.}\ \bibnamefont {Mostepanenko}},\
  }\bibfield  {title} {\enquote {\bibinfo {title} {{New developments in the
  Casimir effect}},}\ }\href {https://doi.org/10.1016/S0370-1573(01)00015-1}
  {\bibfield  {journal} {\bibinfo  {journal} {Phys. Rept.}\ }\textbf {\bibinfo
  {volume} {353}},\ \bibinfo {pages} {1} (\bibinfo {year} {2001})},\ \Eprint
  {http://arxiv.org/abs/quant-ph/0106045} {arXiv:quant-ph/0106045 [quant-ph]}
  \BibitemShut {NoStop}%
\bibitem [{\citenamefont {Milton}(2001)}]{Milton:2001yy}%
  \BibitemOpen
  \bibfield  {author} {\bibinfo {author} {\bibfnamefont {K.~A.}\ \bibnamefont
  {Milton}},\ }\href {https://doi.org/10.1142/4505} {\emph {\bibinfo {title}
  {{The Casimir Effect: Physical Manifestations of Zero-Point Energy}}}}\
  (\bibinfo  {publisher} {World Scientific, Singapore},\ \bibinfo {year}
  {2001})\BibitemShut {NoStop}%
\bibitem [{\citenamefont {Klimchitskaya}\ \emph {et~al.}(2009)\citenamefont
  {Klimchitskaya}, \citenamefont {Mohideen},\ and\ \citenamefont
  {Mostepanenko}}]{Klimchitskaya:2009cw}%
  \BibitemOpen
  \bibfield  {author} {\bibinfo {author} {\bibfnamefont {G.~L.}\ \bibnamefont
  {Klimchitskaya}}, \bibinfo {author} {\bibfnamefont {U.}~\bibnamefont
  {Mohideen}}, \ and\ \bibinfo {author} {\bibfnamefont {V.~M.}\ \bibnamefont
  {Mostepanenko}},\ }\bibfield  {title} {\enquote {\bibinfo {title} {{The
  Casimir force between real materials: Experiment and theory}},}\ }\href
  {https://doi.org/10.1103/RevModPhys.81.1827} {\bibfield  {journal} {\bibinfo
  {journal} {Rev. Mod. Phys.}\ }\textbf {\bibinfo {volume} {81}},\ \bibinfo
  {pages} {1827} (\bibinfo {year} {2009})},\ \Eprint
  {http://arxiv.org/abs/0902.4022} {arXiv:0902.4022 [cond-mat.other]}
  \BibitemShut {NoStop}%
\bibitem [{\citenamefont {Gong}\ \emph {et~al.}(2021)\citenamefont {Gong},
  \citenamefont {Corrado}, \citenamefont {Mahbub}, \citenamefont {Shelden},\
  and\ \citenamefont {Munday}}]{Gong:2020ttb}%
  \BibitemOpen
  \bibfield  {author} {\bibinfo {author} {\bibfnamefont {Tao}\ \bibnamefont
  {Gong}}, \bibinfo {author} {\bibfnamefont {Matthew~R.}\ \bibnamefont
  {Corrado}}, \bibinfo {author} {\bibfnamefont {Ahmed~R.}\ \bibnamefont
  {Mahbub}}, \bibinfo {author} {\bibfnamefont {Calum}\ \bibnamefont {Shelden}},
  \ and\ \bibinfo {author} {\bibfnamefont {Jeremy~N.}\ \bibnamefont {Munday}},\
  }\bibfield  {title} {\enquote {\bibinfo {title} {{Recent progress in
  engineering the Casimir effect \textendash{} Applications to nanophotonics,
  nanomechanics, and chemistry}},}\ }\href
  {https://doi.org/10.1515/nanoph-2020-0425} {\bibfield  {journal} {\bibinfo
  {journal} {Nanophotonics}\ }\textbf {\bibinfo {volume} {10}},\ \bibinfo
  {pages} {523} (\bibinfo {year} {2021})}\BibitemShut {NoStop}%
\bibitem [{\citenamefont {Cougo-Pinto}\ \emph {et~al.}(2001)\citenamefont
  {Cougo-Pinto}, \citenamefont {Farina}, \citenamefont {Mendes},\ and\
  \citenamefont {Tort}}]{Cougo-Pinto:2001}%
  \BibitemOpen
  \bibfield  {author} {\bibinfo {author} {\bibfnamefont {M.~V.}\ \bibnamefont
  {Cougo-Pinto}}, \bibinfo {author} {\bibfnamefont {C.}~\bibnamefont {Farina}},
  \bibinfo {author} {\bibfnamefont {J.~F.~M.}\ \bibnamefont {Mendes}}, \ and\
  \bibinfo {author} {\bibfnamefont {A.~C.}\ \bibnamefont {Tort}},\ }\bibfield
  {title} {\enquote {\bibinfo {title} {{On the non-relativistic Casimir
  effect}},}\ }\href {https://doi.org/10.1590/S0103-97332001000100008}
  {\bibfield  {journal} {\bibinfo  {journal} {Braz. J. Phys.}\ }\textbf
  {\bibinfo {volume} {31}},\ \bibinfo {pages} {1} (\bibinfo {year}
  {2001})}\BibitemShut {NoStop}%
\bibitem [{\citenamefont {Fulling}(2003)}]{Fulling:2003zx}%
  \BibitemOpen
  \bibfield  {author} {\bibinfo {author} {\bibfnamefont {S.~A.}\ \bibnamefont
  {Fulling}},\ }\bibfield  {title} {\enquote {\bibinfo {title} {{Systematics of
  the relationship between vacuum energy calculations and heat-kernel
  coefficients}},}\ }\href {https://doi.org/10.1088/0305-4470/36/24/320}
  {\bibfield  {journal} {\bibinfo  {journal} {J. Phys. A}\ }\textbf {\bibinfo
  {volume} {36}},\ \bibinfo {pages} {6857} (\bibinfo {year} {2003})},\ \Eprint
  {http://arxiv.org/abs/quant-ph/0302117} {arXiv:quant-ph/0302117} \BibitemShut
  {NoStop}%
\bibitem [{\citenamefont {Kolomeisky}\ \emph {et~al.}(2013)\citenamefont
  {Kolomeisky}, \citenamefont {Zaidi}, \citenamefont {Langsjoen},\ and\
  \citenamefont {Straley}}]{Kolomeisky:2013zra}%
  \BibitemOpen
  \bibfield  {author} {\bibinfo {author} {\bibfnamefont {Eugene~B.}\
  \bibnamefont {Kolomeisky}}, \bibinfo {author} {\bibfnamefont {Hussain}\
  \bibnamefont {Zaidi}}, \bibinfo {author} {\bibfnamefont {Luke}\ \bibnamefont
  {Langsjoen}}, \ and\ \bibinfo {author} {\bibfnamefont {Joseph~P.}\
  \bibnamefont {Straley}},\ }\bibfield  {title} {\enquote {\bibinfo {title}
  {{Weyl problem and Casimir effects in spherical shell geometry}},}\ }\href
  {https://doi.org/10.1103/PhysRevA.87.042519} {\bibfield  {journal} {\bibinfo
  {journal} {Phys. Rev. A}\ }\textbf {\bibinfo {volume} {87}},\ \bibinfo
  {pages} {042519} (\bibinfo {year} {2013})},\ \Eprint
  {http://arxiv.org/abs/1110.0421} {arXiv:1110.0421 [cond-mat.stat-mech]}
  \BibitemShut {NoStop}%
\bibitem [{\citenamefont {Ulhoa}\ \emph {et~al.}(2017)\citenamefont {Ulhoa},
  \citenamefont {Santos},\ and\ \citenamefont {Khanna}}]{Ulhoa:2017tsg}%
  \BibitemOpen
  \bibfield  {author} {\bibinfo {author} {\bibfnamefont {S.~C.}\ \bibnamefont
  {Ulhoa}}, \bibinfo {author} {\bibfnamefont {A.~F.}\ \bibnamefont {Santos}}, \
  and\ \bibinfo {author} {\bibfnamefont {Faqir~C.}\ \bibnamefont {Khanna}},\
  }\bibfield  {title} {\enquote {\bibinfo {title} {{Galilean covariance,
  Casimir effect and Stefan\textendash{}Boltzmann law at finite
  temperature}},}\ }\href {https://doi.org/10.1142/S0217751X17500944}
  {\bibfield  {journal} {\bibinfo  {journal} {Int. J. Mod. Phys. A}\ }\textbf
  {\bibinfo {volume} {32}},\ \bibinfo {pages} {1750094} (\bibinfo {year}
  {2017})},\ \Eprint {http://arxiv.org/abs/1711.08510} {arXiv:1711.08510
  [hep-th]} \BibitemShut {NoStop}%
\bibitem [{Note1()}]{Note1}%
  \BibitemOpen
  \bibinfo {note} {Using the zeta function regularization and the dimensional
  regularization, the Casimir energy for a dispersion relation with the order
  of $s$ in the $d+1$ dimensional spacetime and the periodic boundary condition
  is represented as \begin {equation} E_\protect \mathrm {Cas}^{[s]} =2
  \protect \frac {1}{(4\pi )^{(d-1)/2}}\protect \frac {\pi ^{-1/2}\Gamma \left
  ( \protect \frac {d+s}{2} \right ) \zeta (d+s)}{\Gamma (-\protect \frac
  {s}{2})} \protect \frac {2^{d-1+s}}{L^{d-1+s}}. \end {equation} When $s$ is
  an even number, $E_\protect \mathrm {Cas}^{[s]}=0$ because of $\Gamma
  (-\protect \frac {s}{2})$. Also, an alternative interpretation, it is well
  known that the Casimir energy for relativistic fields with a nonzero and
  finite mass is characterized by the modified Bessel function, and its
  infinite-mass limit goes to zero~\cite
  {Hays:1979bc,Ambjorn:1981xw,Plunien:1986ca}.}\BibitemShut {Stop}%
\bibitem [{\citenamefont {Ishikawa}\ \emph {et~al.}(2020)\citenamefont
  {Ishikawa}, \citenamefont {Nakayama},\ and\ \citenamefont
  {Suzuki}}]{Ishikawa:2020ezm}%
  \BibitemOpen
  \bibfield  {author} {\bibinfo {author} {\bibfnamefont {Tsutomu}\ \bibnamefont
  {Ishikawa}}, \bibinfo {author} {\bibfnamefont {Katsumasa}\ \bibnamefont
  {Nakayama}}, \ and\ \bibinfo {author} {\bibfnamefont {Kei}\ \bibnamefont
  {Suzuki}},\ }\bibfield  {title} {\enquote {\bibinfo {title} {{Casimir effect
  for lattice fermions}},}\ }\href
  {https://doi.org/10.1016/j.physletb.2020.135713} {\bibfield  {journal}
  {\bibinfo  {journal} {Phys. Lett. B}\ }\textbf {\bibinfo {volume} {809}},\
  \bibinfo {pages} {135713} (\bibinfo {year} {2020})},\ \Eprint
  {http://arxiv.org/abs/2005.10758} {arXiv:2005.10758 [hep-lat]} \BibitemShut
  {NoStop}%
\bibitem [{\citenamefont {Ishikawa}\ \emph {et~al.}(2021)\citenamefont
  {Ishikawa}, \citenamefont {Nakayama},\ and\ \citenamefont
  {Suzuki}}]{Ishikawa:2020icy}%
  \BibitemOpen
  \bibfield  {author} {\bibinfo {author} {\bibfnamefont {Tsutomu}\ \bibnamefont
  {Ishikawa}}, \bibinfo {author} {\bibfnamefont {Katsumasa}\ \bibnamefont
  {Nakayama}}, \ and\ \bibinfo {author} {\bibfnamefont {Kei}\ \bibnamefont
  {Suzuki}},\ }\bibfield  {title} {\enquote {\bibinfo {title}
  {{Lattice-fermionic Casimir effect and topological insulators}},}\ }\href
  {https://doi.org/10.1103/PhysRevResearch.3.023201} {\bibfield  {journal}
  {\bibinfo  {journal} {Phys. Rev. Research}\ }\textbf {\bibinfo {volume}
  {3}},\ \bibinfo {pages} {023201} (\bibinfo {year} {2021})},\ \Eprint
  {http://arxiv.org/abs/2012.11398} {arXiv:2012.11398 [hep-lat]} \BibitemShut
  {NoStop}%
\bibitem [{Note2()}]{Note2}%
  \BibitemOpen
  \bibinfo {note} {Studies of the Casimir effect using the lattice
  regularization are still few. For early works for scalar fields, see
  Refs.~\cite {Actor:1999nb,Pawellek:2013sda}.}\BibitemShut {Stop}%
\bibitem [{Note3()}]{Note3}%
  \BibitemOpen
  \bibinfo {note} {In Supplemental Material S1, we compare the definitions of
  Casimir energies in the continuum space and in the lattice space. In S2, we
  show the detailed analyses of the phononic Casimir effect in a
  one-dimensional GaAs nanowire.}\BibitemShut {Stop}%
\bibitem [{\citenamefont {Hays}(1979)}]{Hays:1979bc}%
  \BibitemOpen
  \bibfield  {author} {\bibinfo {author} {\bibfnamefont {P.}~\bibnamefont
  {Hays}},\ }\bibfield  {title} {\enquote {\bibinfo {title} {{Vacuum
  fluctuations of a confined massive field in two dimensions}},}\ }\href
  {https://doi.org/10.1016/0003-4916(79)90090-3} {\bibfield  {journal}
  {\bibinfo  {journal} {Annals Phys.}\ }\textbf {\bibinfo {volume} {121}},\
  \bibinfo {pages} {32} (\bibinfo {year} {1979})}\BibitemShut {NoStop}%
\bibitem [{\citenamefont {Ambj\o{}rn}\ and\ \citenamefont
  {Wolfram}(1983)}]{Ambjorn:1981xw}%
  \BibitemOpen
  \bibfield  {author} {\bibinfo {author} {\bibfnamefont {Jan}\ \bibnamefont
  {Ambj\o{}rn}}\ and\ \bibinfo {author} {\bibfnamefont {Stephen}\ \bibnamefont
  {Wolfram}},\ }\bibfield  {title} {\enquote {\bibinfo {title} {{Properties of
  the vacuum. I. Mechanical and thermodynamic}},}\ }\href
  {https://doi.org/10.1016/0003-4916(83)90065-9} {\bibfield  {journal}
  {\bibinfo  {journal} {Annals Phys.}\ }\textbf {\bibinfo {volume} {147}},\
  \bibinfo {pages} {1} (\bibinfo {year} {1983})}\BibitemShut {NoStop}%
\bibitem [{Note4()}]{Note4}%
  \BibitemOpen
  \bibinfo {note} {By using the dispersion relation $a\protect \tilde {\protect
  \bf k}$ on the lattice, the Taylor expansion of Eq.~(\ref {eq:E0_exp})
  converges when $\DOTSB \sum@ \slimits@ _{i} (2-2\protect \qopname \relax
  o{cos}ak_i)/(am)^2<1$}\BibitemShut {NoStop}%
\bibitem [{\citenamefont {McCann}\ and\ \citenamefont
  {Fal'ko}(2006)}]{McCann:2005xf}%
  \BibitemOpen
  \bibfield  {author} {\bibinfo {author} {\bibfnamefont {Edward}\ \bibnamefont
  {McCann}}\ and\ \bibinfo {author} {\bibfnamefont {Vladimir~I.}\ \bibnamefont
  {Fal'ko}},\ }\bibfield  {title} {\enquote {\bibinfo {title} {{Landau-Level
  Degeneracy and Quantum Hall Effect in a Graphite Bilayer}},}\ }\href
  {https://doi.org/10.1103/PhysRevLett.96.086805} {\bibfield  {journal}
  {\bibinfo  {journal} {Phys. Rev. Lett.}\ }\textbf {\bibinfo {volume} {96}},\
  \bibinfo {pages} {086805} (\bibinfo {year} {2006})},\ \Eprint
  {http://arxiv.org/abs/cond-mat/0510237} {arXiv:cond-mat/0510237} \BibitemShut
  {NoStop}%
\bibitem [{\citenamefont {McCann}\ and\ \citenamefont
  {Koshino}(2013)}]{McCann:2013}%
  \BibitemOpen
  \bibfield  {author} {\bibinfo {author} {\bibfnamefont {Edward}\ \bibnamefont
  {McCann}}\ and\ \bibinfo {author} {\bibfnamefont {Mikito}\ \bibnamefont
  {Koshino}},\ }\bibfield  {title} {\enquote {\bibinfo {title} {The electronic
  properties of bilayer graphene},}\ }\href
  {https://doi.org/10.1088/0034-4885/76/5/056503} {\bibfield  {journal}
  {\bibinfo  {journal} {Rep. Prog. Phys.}\ }\textbf {\bibinfo {volume} {76}},\
  \bibinfo {pages} {056503} (\bibinfo {year} {2013})},\ \Eprint
  {http://arxiv.org/abs/1205.6953} {arXiv:1205.6953 [cond-mat.mes-hall]}
  \BibitemShut {NoStop}%
\bibitem [{\citenamefont {Xu}\ \emph {et~al.}(2011)\citenamefont {Xu},
  \citenamefont {Weng}, \citenamefont {Wang}, \citenamefont {Dai},\ and\
  \citenamefont {Fang}}]{Xu:2011dn}%
  \BibitemOpen
  \bibfield  {author} {\bibinfo {author} {\bibfnamefont {Gang}\ \bibnamefont
  {Xu}}, \bibinfo {author} {\bibfnamefont {Hongming}\ \bibnamefont {Weng}},
  \bibinfo {author} {\bibfnamefont {Zhijun}\ \bibnamefont {Wang}}, \bibinfo
  {author} {\bibfnamefont {Xi}~\bibnamefont {Dai}}, \ and\ \bibinfo {author}
  {\bibfnamefont {Zhong}\ \bibnamefont {Fang}},\ }\bibfield  {title} {\enquote
  {\bibinfo {title} {{Chern Semimetal and the Quantized Anomalous Hall Effect
  in ${\mathrm{HgCr}}_{2}{\mathrm{Se}}_{4}$}},}\ }\href
  {https://doi.org/10.1103/PhysRevLett.107.186806} {\bibfield  {journal}
  {\bibinfo  {journal} {Phys. Rev. Lett.}\ }\textbf {\bibinfo {volume} {107}},\
  \bibinfo {pages} {186806} (\bibinfo {year} {2011})},\ \Eprint
  {http://arxiv.org/abs/1106.3125} {arXiv:1106.3125 [cond-mat.mes-hall]}
  \BibitemShut {NoStop}%
\bibitem [{\citenamefont {Fang}\ \emph {et~al.}(2012)\citenamefont {Fang},
  \citenamefont {Gilbert}, \citenamefont {Dai},\ and\ \citenamefont
  {Bernevig}}]{Fang:2012}%
  \BibitemOpen
  \bibfield  {author} {\bibinfo {author} {\bibfnamefont {Chen}\ \bibnamefont
  {Fang}}, \bibinfo {author} {\bibfnamefont {Matthew~J.}\ \bibnamefont
  {Gilbert}}, \bibinfo {author} {\bibfnamefont {Xi}~\bibnamefont {Dai}}, \ and\
  \bibinfo {author} {\bibfnamefont {B.~Andrei}\ \bibnamefont {Bernevig}},\
  }\bibfield  {title} {\enquote {\bibinfo {title} {{Multi-Weyl Topological
  Semimetals Stabilized by Point Group Symmetry}},}\ }\href
  {https://doi.org/10.1103/PhysRevLett.108.266802} {\bibfield  {journal}
  {\bibinfo  {journal} {Phys. Rev. Lett.}\ }\textbf {\bibinfo {volume} {108}},\
  \bibinfo {pages} {266802} (\bibinfo {year} {2012})},\ \Eprint
  {http://arxiv.org/abs/1111.7309} {arXiv:1111.7309 [cond-mat.mes-hall]}
  \BibitemShut {NoStop}%
\bibitem [{\citenamefont {Huang}\ \emph {et~al.}(2016)\citenamefont {Huang},
  \citenamefont {Xu}, \citenamefont {Belopolski}, \citenamefont {Lee},
  \citenamefont {Chang}, \citenamefont {Chang}, \citenamefont {Wang},
  \citenamefont {Alidoust}, \citenamefont {Bian},\ and\ \citenamefont {{\it et
  al.}}}]{Huang:2016}%
  \BibitemOpen
  \bibfield  {author} {\bibinfo {author} {\bibfnamefont {Shin-Ming}\
  \bibnamefont {Huang}}, \bibinfo {author} {\bibfnamefont {Su-Yang}\
  \bibnamefont {Xu}}, \bibinfo {author} {\bibfnamefont {Ilya}\ \bibnamefont
  {Belopolski}}, \bibinfo {author} {\bibfnamefont {Chi-Cheng}\ \bibnamefont
  {Lee}}, \bibinfo {author} {\bibfnamefont {Guoqing}\ \bibnamefont {Chang}},
  \bibinfo {author} {\bibfnamefont {Tay-Rong}\ \bibnamefont {Chang}}, \bibinfo
  {author} {\bibfnamefont {BaoKai}\ \bibnamefont {Wang}}, \bibinfo {author}
  {\bibfnamefont {Nasser}\ \bibnamefont {Alidoust}}, \bibinfo {author}
  {\bibfnamefont {Guang}\ \bibnamefont {Bian}}, \ and\ \bibinfo {author}
  {\bibfnamefont {Madhab~Neupane}\ \bibnamefont {{\it et al.}}},\ }\bibfield
  {title} {\enquote {\bibinfo {title} {{New type of Weyl semimetal with
  quadratic double Weyl fermions}},}\ }\href
  {https://doi.org/10.1073/pnas.1514581113} {\bibfield  {journal} {\bibinfo
  {journal} {Proc. Natl. Acad. Sci. U.S.A.}\ }\textbf {\bibinfo {volume}
  {113}},\ \bibinfo {pages} {1180} (\bibinfo {year} {2016})},\ \Eprint
  {http://arxiv.org/abs/1503.05868} {arXiv:1503.05868 [cond-mat.mes-hall]}
  \BibitemShut {NoStop}%
\bibitem [{\citenamefont {Owerre}(2016)}]{Owerre:2016}%
  \BibitemOpen
  \bibfield  {author} {\bibinfo {author} {\bibfnamefont {S.~A.}\ \bibnamefont
  {Owerre}},\ }\bibfield  {title} {\enquote {\bibinfo {title} {{Magnon Hall
  effect in AB-stacked bilayer honeycomb quantum magnets}},}\ }\href
  {https://doi.org/10.1103/PhysRevB.94.094405} {\bibfield  {journal} {\bibinfo
  {journal} {Phys. Rev. B}\ }\textbf {\bibinfo {volume} {94}},\ \bibinfo
  {pages} {094405} (\bibinfo {year} {2016})},\ \Eprint
  {http://arxiv.org/abs/1604.05292} {arXiv:1604.05292 [cond-mat.str-el]}
  \BibitemShut {NoStop}%
\bibitem [{\citenamefont {Strauch}\ and\ \citenamefont
  {Dorner}(1990)}]{Strauch:1990}%
  \BibitemOpen
  \bibfield  {author} {\bibinfo {author} {\bibfnamefont {D.}~\bibnamefont
  {Strauch}}\ and\ \bibinfo {author} {\bibfnamefont {B.}~\bibnamefont
  {Dorner}},\ }\bibfield  {title} {\enquote {\bibinfo {title} {{Phonon
  dispersion in GaAs}},}\ }\href {https://doi.org/10.1088/0953-8984/2/6/006}
  {\bibfield  {journal} {\bibinfo  {journal} {J. Phys. Condens. Matter}\
  }\textbf {\bibinfo {volume} {2}},\ \bibinfo {pages} {1457} (\bibinfo {year}
  {1990})}\BibitemShut {NoStop}%
\bibitem [{\citenamefont {Actor}\ \emph {et~al.}(2000)\citenamefont {Actor},
  \citenamefont {Bender},\ and\ \citenamefont {Reingruber}}]{Actor:1999nb}%
  \BibitemOpen
  \bibfield  {author} {\bibinfo {author} {\bibfnamefont {A.}~\bibnamefont
  {Actor}}, \bibinfo {author} {\bibfnamefont {I.}~\bibnamefont {Bender}}, \
  and\ \bibinfo {author} {\bibfnamefont {J.}~\bibnamefont {Reingruber}},\
  }\bibfield  {title} {\enquote {\bibinfo {title} {{Casimir effect on a finite
  lattice}},}\ }\href
  {https://doi.org/10.1002/(SICI)1521-3978(200004)48:4<303::AID-PROP303>3.0.CO;2-J}
  {\bibfield  {journal} {\bibinfo  {journal} {Fortsch. Phys.}\ }\textbf
  {\bibinfo {volume} {48}},\ \bibinfo {pages} {303} (\bibinfo {year} {2000})},\
  \Eprint {http://arxiv.org/abs/quant-ph/9908058} {arXiv:quant-ph/9908058
  [quant-ph]} \BibitemShut {NoStop}%
\bibitem [{\citenamefont {Pawellek}()}]{Pawellek:2013sda}%
  \BibitemOpen
  \bibfield  {author} {\bibinfo {author} {\bibfnamefont {Michael}\ \bibnamefont
  {Pawellek}},\ }\bibfield  {title} {\enquote {\bibinfo {title} {{Finite-sites
  corrections to the Casimir energy on a periodic lattice}},}\ }\href@noop {}
  {\ }\Eprint {http://arxiv.org/abs/1303.4708} {arXiv:1303.4708 [hep-th]}
  \BibitemShut {NoStop}%
\bibitem [{\citenamefont {Ashcroft}\ and\ \citenamefont
  {Mermin}(1976)}]{Ashcroft:1976}%
  \BibitemOpen
  \bibfield  {author} {\bibinfo {author} {\bibfnamefont {N.~W.}\ \bibnamefont
  {Ashcroft}}\ and\ \bibinfo {author} {\bibfnamefont {N.~D.}\ \bibnamefont
  {Mermin}},\ }\href@noop {} {\emph {\bibinfo {title} {{Solid State
  Physics}}}}\ (\bibinfo  {publisher} {Saunders College Publishing, New York},\
  \bibinfo {year} {1976})\BibitemShut {NoStop}%
\end{thebibliography}%

\clearpage
\widetext
\noindent
{\fontsize{11pt}{11pt}\selectfont \bf Supplemental Material for ``Remnants of the nonrelativistic Casimir effect on the lattice"}

  \setcounter{section}{0}
  \setcounter{equation}{0}
  \setcounter{figure}{0}
  \renewcommand{\theequation}{S\arabic{equation}}
  \renewcommand{\thesection}{S\arabic{section}}
  \renewcommand{\thefigure}{S\arabic{figure}}

\section{Definitions of Casimir energies in continuum space and in lattice space}
In the main text, we consider the Casimir effect defined on the lattice, whereas the conventional photonic Casimir effect is defined in the continuous spacetime.
In this Supplemental Material, we compare these two definitions of the Casimir effects in continuous spacetime and in lattice space.

\subsection{In continuous spacetime}
Following the definition by Casimir in the original paper~\cite{Casimir:1948dh}, the Casimir energy for photon fields in the $3+1$ dimensional continuous spacetime is defined as 
\begin{subequations}
\begin{align}
E_\mathrm{Cas}^\mathrm{cont} &\equiv E_0^\mathrm{cont,sum}(L_z) - E_0^\mathrm{cont,int}(L_z), \\
E_0^\mathrm{cont,sum}(L_z) & = \hbar c L_xL_y \sum_j \int_{-\infty}^\infty \int_{-\infty}^\infty \frac{dk_x}{2\pi} \frac{dk_y}{2\pi} \left[ \frac{1}{2} \sum_{n=(0)1}^{\infty} \sqrt{k_x^2+k_y^2 + \frac{n^2\pi^2}{L_z^2}}  \right], \\
E_0^\mathrm{cont,int}(L_z) & = \hbar c L_xL_y \sum_j \int_{-\infty}^\infty \int_{-\infty}^\infty \int_{-\infty}^\infty \frac{d^3 \bf{k}}{(2\pi)^3} \left[ \frac{L_z}{2} \sqrt{k_x^2+k_y^2 + k_z^2} \right].
\end{align}
\end{subequations}
$\sum_j$ is the sum of the number of degrees of freedom  (the two polarizations for photons).
$\sum_{n=(0)1}$ means that we have to multiply the term with $n=0$ by an additional factor of $1/2$.
Important quantities are two types of the zero-point energies, $E_0^\mathrm{cont,sum}$ and $E_0^\mathrm{cont,int}$:
\begin{enumerate}
\item $E_0^\mathrm{cont,sum}$ is the zero-point energy for photon fields in the space sandwiched by two parallel plates (in other words, photon fields {\it in a finite $L_z$}).
Due to the existence of the boundary conditions, the $z$-momentum and eigenvalues are discretized and labeled by $n=0, 1, \cdots, \infty$.
\item $E_0^\mathrm{cont,int}$ is the zero-point energy for photon fields {\it in the infinite $L_z$}.
The spatial momenta and the energy eigenvalues are continuous.
We remark that this term is proportional to $L_z$, which means that $E_0^\mathrm{cont,int}$ is the zero-point energy for a volume $V=L_x L_y L_z$ with a small length $L_z$.
\item Both $E_0^\mathrm{cont,sum}$ and $E_0^\mathrm{cont,int}$ are {\it divergent} due to the infinite momentum integral and the infinite momentum sum.
Hence, in order to obtain a finite Casimir energy, one has to apply a mathematical regularization technique, such as the Euler-Maclaurin formula and the zeta function regularization.
\end{enumerate}

\subsection{In lattice space}
Similar to the definition in the continuous spacetime, the counterpart in the lattice space with the lattice spacing $a$ is as follows~\cite{Actor:1999nb,Pawellek:2013sda,Ishikawa:2020ezm,Ishikawa:2020icy}:
Using the replacements such as $L_x dk_x \to N_x d(ak_x)$,
\begin{subequations}
\begin{align}
E_\mathrm{Cas}^\mathrm{lat} &\equiv E_0^\mathrm{lat,sum}(N_z) - E_0^\mathrm{lat,int}(N_z), \\
E_0^\mathrm{lat,sum}(N_z) &= \hbar cN_x N_y \sum_j \int_\mathrm{BZ} \frac{d(ak_x)}{2\pi} \frac{d(ak_y)}{2\pi} \left[ \frac{1}{2} \sum_{n}^\mathrm{BZ} |\omega_{k_\perp,n,j} |  \right], \\
E_0^\mathrm{lat,int}(N_z) & =  \hbar c N_x N_y \sum_j \int_\mathrm{BZ}  \frac{d(ak_x)}{2\pi} \frac{d(ak_y)}{2\pi}  \left[ \frac{N_z}{2} \int_\mathrm{BZ} \frac{d(ak_z)}{2\pi} |\omega_{{\bf k},j} | \right],
\end{align}
\end{subequations}
where the integrals and the sum are taken within the first Brillouin zone (BZ).

We note some comments: (i) In the main text, we define $E_\mathrm{Cas} \equiv E_\mathrm{Cas}^\mathrm{lat} /N_xN_y$, which means the Casimir energy per ``surface area" $N_x N_y$.
In this sense, $E_\mathrm{Cas}$ may be called ``energy surface density" with the dimension of the energy.
(ii) In the main text, we use natural units $\hbar = c=1$.
(iii) In the main text, for simplicity, we use dimensionless quantities with the lattice spacing, such as $a E_\mathrm{Cas}$ and $a \omega_{\bf k}$.
(iv) In the main text, the number of degrees of freedom, labeled by $j$, is fixed as $1$ because we consider the real scalar field.
If there are particle/antiparticle and spin degrees of freedom, we need to sum all the possible modes.

$E_\mathrm{Cas}$, $E_0^\mathrm{int}$, and $E_0^\mathrm{sum}$ used in the main text are the same as $E_\mathrm{Cas}^\mathrm{lat}$, $E_0^\mathrm{lat,int}$, $E_0^\mathrm{lat,sum}$, except for the factor of $N_x N_y$.
We remark that $E_0^\mathrm{int}$ and $E_0^\mathrm{sum}$ are {\it not divergent} because of the lattice regularization.
Therefore, the three quantities of $E_0^\mathrm{int}$, $E_0^\mathrm{sum}$, and $E_\mathrm{Cas}$ are well-defined observables.
In experiments, $E_0^\mathrm{int}$ is observed as the zero-point energy (the free energy at zero temperature) when a material is large enough.
$E_0^\mathrm{sum}$ is observed as the zero-point energy when the $z$ direction of a material is short enough.
$E_\mathrm{Cas}$ is obtained as the difference between $E_0^\mathrm{int}$ and $E_0^\mathrm{sum}$.
Thus, on the lattice, $E_0^\mathrm{int}$ and $E_0^\mathrm{sum}$ are {\it automatically} regularized, which is different from the definition in the continuous spacetime.
The quantity of $E_\mathrm{Cas}$ is similar to each other.

\subsection{Physical setup}
\begin{figure}[h!]
    \centering
    \begin{minipage}[t]{0.5\columnwidth}
    \includegraphics[clip,width=1.0\columnwidth]{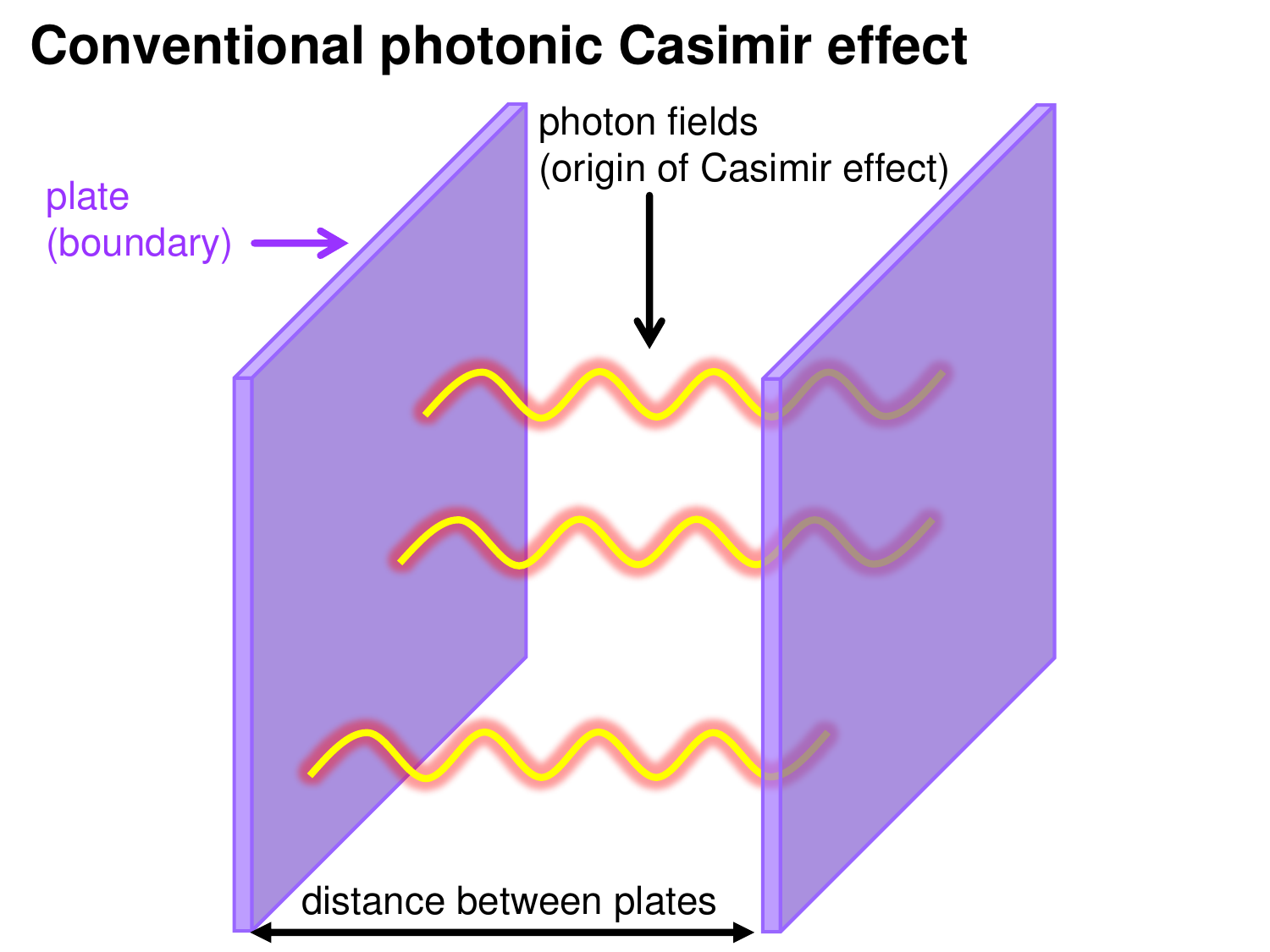}
    \end{minipage}%
    \begin{minipage}[t]{0.5\columnwidth}
    \includegraphics[clip,width=1.0\columnwidth]{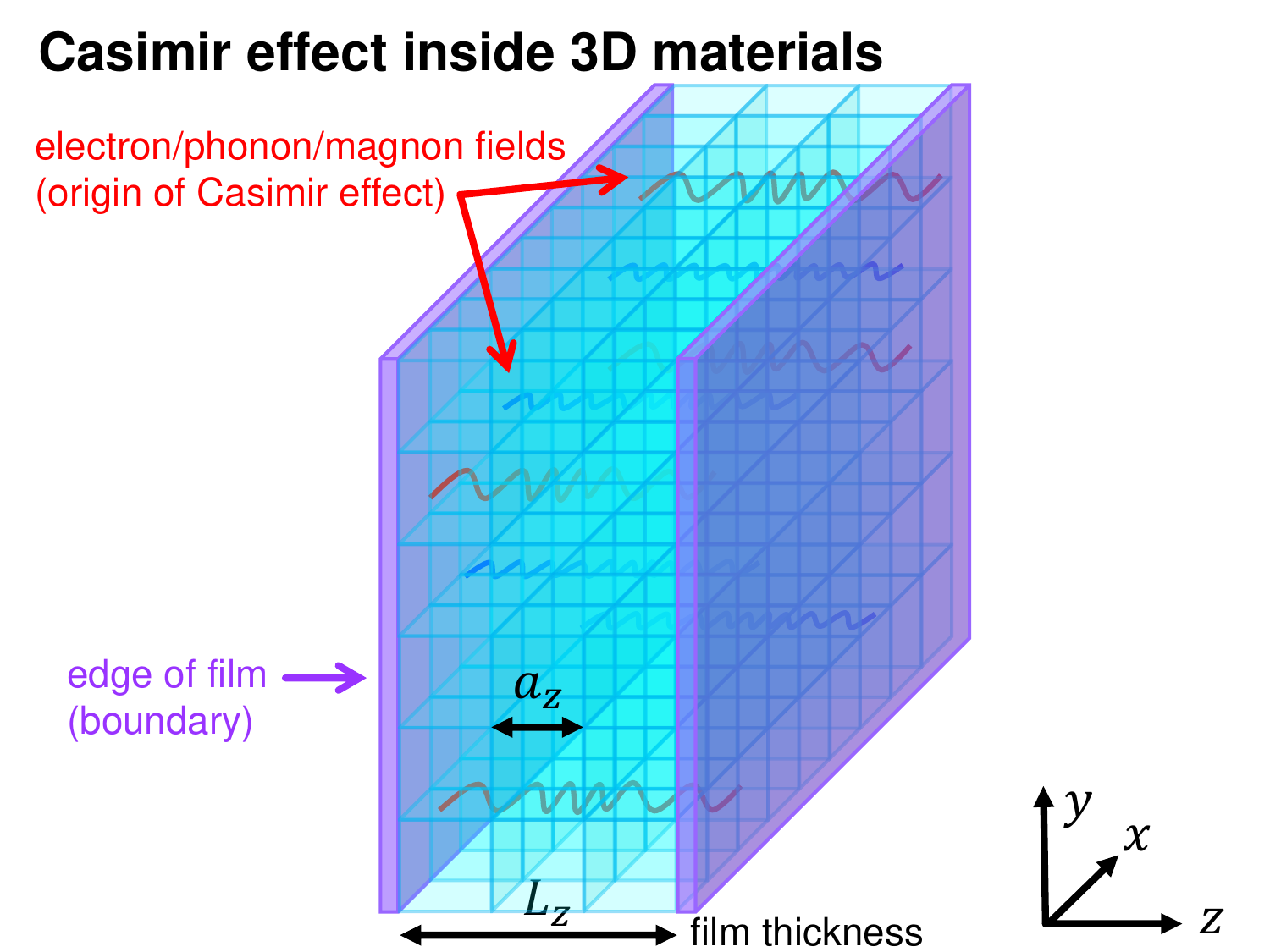}
    \end{minipage}
    \caption{Left: Conventional photonic Casimir effect. Right: Casimir effect inside 3D materials.}
    \label{fig:conv-lat}
\end{figure}

Finally, we explain a physical setup experimentally realizing the Casimir effect on the lattice.
In Fig.~\ref{fig:conv-lat}, we compare the conventional photonic Casimir effect between two parallel plates and Casimir effects for quantum fields (on the lattice) inside a thin film of three-dimensional (3D) materials.
Inside materials, quantum fields, such as electrons, phonons, and magnons, are in lattice space.
The boundary conditions of the two parallel plates in the conventional Casimir effect correspond to those of the edges of the film in the $z$ direction.

While Fig.~\ref{fig:conv-lat} shows the 3D geometry of thin films, the lower dimensions are also realized as narrow nanoribbons of 2D materials and short nanochains of 1D materials.
Furthermore, the geometry with the periodic boundary condition can be realized as a solid torus structure of 3D materials and the winding direction of a carbon nanotube.

\section{Phononic Casimir effect in 1D GaAs nanowire}
Here, we demonstrate an approximate realization of the remnant Casimir effect in a realistic material.
In particular, we focus on the {\it phononic Casimir effect} originating from phonon fields in  gallium arsenide (GaAs) which is a semiconductor.
Note that the phonon fields are treated as real scalar fields on the lattice, and one can derive the existence of the zero-point energy.
We consider a nanowire of GaAs along the $z$ axis, where the transverse momenta, $k_x$ and $k_y$, are dominated by only the lowest energy modes discretized by the finite-size effect, and then we can regard this situation as an effective one-dimensional system along the $z$ axis.
In this material, there exist six phonons: two transverse acoustic (TA) modes, one longitudinal acoustic (LA) mode, two transverse optical (TO) modes, and one longitudinal optical (LO) mode, which can move along the $z$ axis.

\begin{figure}[b!]
    \centering
    \begin{minipage}[t]{0.5\columnwidth}
    \includegraphics[clip,width=1.0\columnwidth]{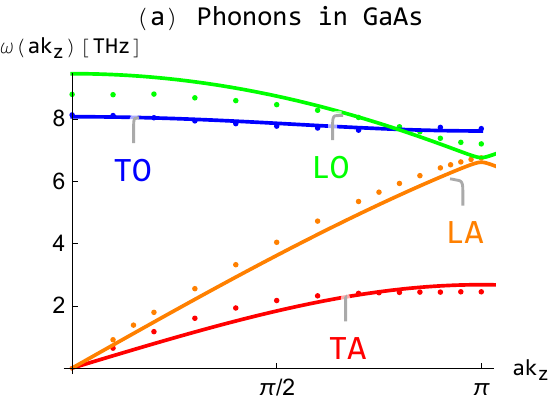}
    \end{minipage}%
    \begin{minipage}[t]{0.5\columnwidth}
    \includegraphics[clip,width=1.0\columnwidth]{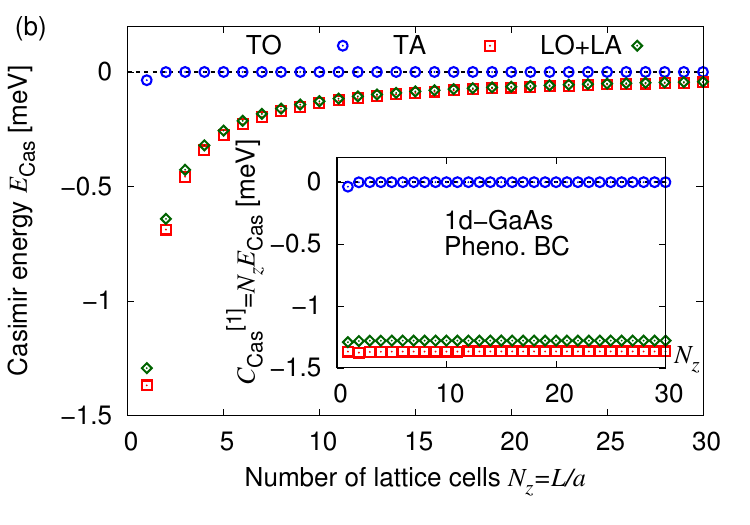}
    \end{minipage}
    \caption{(a) Phonon dispersion relations in $k_z$ space in GaAs.
The curves are obtained from our model, and the points are input data from experiments~\cite{Strauch:1990}.
(b) Casimir energy for phonon fields in one-dimensional-like GaAs nanowires {\it with a phenomenological boundary condition}.
}
    \label{fig:GaAs_phenoBC}
\end{figure}

In order to construct a model of the dispersion relations of these phonons, we apply the well-known formula for a one-dimensional chain with two types of atoms (with a force constant $C$ and masses $M_1$ and $M_2$)~\cite{Ashcroft:1976}:
\begin{align}
\omega_\pm^2 (k_z) = \frac{C(M_1+M_2) \pm  C\sqrt{M_1^2+M_2^2+2M_1M_2\cos{ak_z} } }{M_1M_2}, \label{1Dphonon}
\end{align}
where $+$ and $-$ signs correspond to the optical and acoustic phonons, respectively.
For the phonons in GaAs, the dispersions were measured by inelastic neutron-scattering experiment~\cite{Strauch:1990}.
By fitting the experimental data in the $k_z$-directed Brillouin zone from ${\bf k}=0$, we determine the values of the three parameters, $C,M_1$, and $M_2$.
For the TA and TO modes, $C \sim 120.74$ THz, $M_1 \sim 33.70$ THz$^{-1}$, and $M_2 \sim 4.16$ THz$^{-1}$.
For the LA and LO modes, $C \sim 88.39$ THz, $M_1 \sim 4.05$ THz$^{-1}$, and $M_2 \sim 3.87$ THz$^{-1}$.
In Fig.~\ref{fig:GaAs_phenoBC}(a), we compare the theoretical dispersion curves obtained from these parameters and the experimental data~\cite{Strauch:1990}.
Note that Eq.~(\ref{1Dphonon}) should be regarded as a rough model with only the three parameters, but it is enough to reproduce the experimental phonon spectrum (especially, for the TO modes which we focus on below).

In this setup, when boundary conditions are introduced, $k_z$ is discretized, and hence the Casimir effect for phonon fields arises.
We calculate the Casimir energy by using the one-dimensional form of the approach used in the main text.
The numerical results for the periodic boundary condition (which may be regarded as a ring-like geometry) are shown in the main text.
In this Supplemental Material, we show the result with a phenomenological boundary condition (which can be regarded as a short-wire geometry), where the momentum is discretized as $ak_z \to l \pi /N_z$ with $l=1,\cdots,2N_z$ (or equivalently $l=0,\cdots,2N_z-1)$, and the sum is over $\frac{1}{2}\sum_{l=1}^{2N_z}$ (the factor of $\frac{1}{2}$ is required because of the number of eigenvalues).
Note that, for the quadratic dispersion, a qualitative difference between the periodic and phenomenological boundaries is the behavior at $N_z=1$, but the behavior at $N_z \geq 2$ is similar to each other,

In Fig.~\ref{fig:GaAs_phenoBC}(b), we show the numerical results.
In the following, we summarize the relationship between the four types of phonons and the behaviors of Casimir effects:
\begin{itemize}
\item TO modes: When the dispersions around $ak_z=0$ and $\pi$ are expanded in powers of $(ak_z)^2$ and  $(ak_z-\pi)^2$, respectively, these are regarded as approximately quadratic dispersions. 
As a result, the remnant Casimir effect is approximately realized, and the Casimir energy is strongly suppressed in a long distance. 
\item TA modes: The dispersion around $ak_z=0$ is linear, while the dispersion around $ak_z=\pi$ is approximately quadratic.
As a result, the lasting Casimir effect is realized by the linear band structure around $ak_z=0$. 
\item LO mode: The dispersion around $ak_z=0$ is approximately quadratic, while that around $ak_z=\pi$ is approximately linear.
Thus, although the functional forms~(\ref{1Dphonon}) of the TO and LO modes are the same, the choice of different model parameters changes the band structures around $ak_z=\pi$, and, as a result, the typical behavior of the Casimir effect is distinct.
\item LA mode: The dispersion around $ak_z=0$ is linear, which is similar to that of the TA modes. 
The dispersion around $ak_z=\pi$ is approximately linear, which is similar to that of the LO modes.
\end{itemize}

Note that the dispersion relation around $ak_z=\pi$ can be characterized by the two model parameters, $M_1$ and $M_2$ ($C$ is just an overall factor).
When we expand the right-hand side of Eq.~(\ref{1Dphonon}) around $ak_z=\pi$ and substitute the model parameters, we obtain 
\begin{align}
&\omega_\mathrm{TO} (k_z) = 7.62 + 0.13 (ak_z-\pi)^2 - 1.8 \times 10^{-2} (ak_z-\pi)^4 + 2.0 \times 10^{-3} (ak_z-\pi)^6 +\cdots \nonumber\\
&\omega_\mathrm{TA} (k_z) = 2.68 - 0.38  (ak_z-\pi)^2 + 2.0 \times 10^{-2} (ak_z-\pi)^4 - 2.0 \times 10^{-3} (ak_z-\pi)^6 +\cdots \nonumber\\
&\omega_\mathrm{LO} (k_z) = 6.76 + 18  (ak_z-\pi)^2 - 2.3 \times 10^3 (ak_z-\pi)^4 + 5.7 \times 10^{5} (ak_z-\pi)^6 +\cdots \nonumber\\
&\omega_\mathrm{LA} (k_z) = 6.61 - 19  (ak_z-\pi)^2 + 2.3 \times 10^3 (ak_z-\pi)^4 + 5.7 \times 10^{5} (ak_z-\pi)^6 +\cdots, \nonumber
\end{align}
where we keep two or three significant digits of the coefficients (in the unit of THz).
From these coefficients, we can regard the dispersions of the TO and TA modes around $ak_z=\pi$ as quadratic-like ones, while those of the LO and LA modes are not quadratic.
\end{document}